# Memory Effects in the Charge Response of Lightly Doped La$_{2-x}$Sr$_x$CuO$_4$

I. Raičević, Dragana Popović*, C. Panagopoulos, T. Sasagawa

**Abstract** The in-plane magnetoresistance (MR) of a single crystal La$_{1.97}$Sr$_{0.03}$CuO$_4$ has been studied at low temperatures $T$ using several experimental protocols. At $T$ well below the spin-glass transition temperature, the MR becomes positive and exhibits several glassy features, such as history dependence, memory and hysteresis. These observations are qualitatively similar to the previously reported behavior of the out-of-plane resistance. The results suggest that the memory effects in the MR are related to the onset of glassiness in the dynamics of doped holes.

## 1. Introduction

The complex interplay of spin, charge and superconducting orders is believed to lead to various nanoscale inhomogeneities in underdoped cuprates and possibly provide an explanation for many surprising phenomena observed in high-temperature superconductors (HTSs). In general, many different arrangements of such nanoscopic ordered regions often have comparable energies, resulting in slow, complex dynamics typical of glassy systems. Indeed, in cuprates, spin-glass (SG) behavior is well established at temperatures $T<T_{SG}(x)$ (Fig. 1 inset), where $x$ is the doping [1-5]. Although studies of charge dynamics have been, by comparison, relatively scarce, recent experiments on La$_{2-x}$Sr$_x$CuO$_4$ (LSCO), the prototypical cuprate HTS, have provided evidence for the glassy freezing of charges as $T\rightarrow 0$.

In particular, in LSCO with $x$=0.03, which is located in the pseudogap regime, but it does not superconduct at any $T$ (Fig. 1 inset), resistance noise spectroscopy [6, 7] shows that, deep within the SG phase at $T \ll T_{SG}$, doped holes form a collective, glassy state of charge domains or clusters located in the CuO$_2$ planes. These conclusions are supported by impedance spectroscopy [8]. In the same $T$ range, both out-of-plane and in-plane magnetoresistance (MR) exhibit the emergence of a strong, positive component for all orientations of the magnetic field $B$

I. Raičević, D. Popović*
National High Magnetic Field Laboratory, Florida State University, Tallahassee, FL 32310, USA
* Corresponding author. Tel.: +1-850-644-3913; fax: +1-850-644-5038; e-mail: dragana@magnet.fsu.edu

C. Panagopoulos
Division of Physics and Applied Physics, Nanyang Technological University, Singapore
Department of Physics and FORTH, University of Crete, Greece

T. Sasagawa
Materials and Structures Laboratory, Tokyo Institute of Technology, Kanagawa, Japan

[6, 9]. Surprisingly, the mechanism of this positive MR (pMR) appears to be the same as that observed in various nonmagnetic, disordered materials with strong Coulomb interactions. Namely, in a system that conducts via variable-range hopping (VRH), the Zeeman splitting in the presence of a Coulomb repulsion between two holes in the same disorder-localized state leads to a pMR by blocking certain hopping channels [10, 11]. The data are consistent with the picture in which correlated charge clusters coexist with hole-poor antiferromagnetic (AF) domains that, to leading order, remain frozen at very low $T \ll T_{SG}$, at least on experimental time scales. Therefore, only holes in the domain "walls" (i.e. patchy areas separating the AF domains) contribute to transport and glassiness in the resistance noise.

The pMR grows rapidly with decreasing $T$ and thus dominates the MR in the entire experimental $B$ range at the lowest $T$. At higher $T$ and $B$, on the other hand, the MR is negative. The mechanism of the negative MR at high $T$ is attributed [12] to the reorientation of the weak ferromagnetic (FM) moments [13, 14] associated with each AF domain and oriented along the $c$ axis (perpendicular to $CuO_2$ planes). Unlike the negative MR, the pMR reveals clear signatures of glassiness, such as hysteresis and memory [6, 9]. Here we present a more detailed study of the intriguing memory effects in the in-plane transport with $B$ applied parallel to $CuO_2$ (ab) planes.

## 2. Experiment

A high quality single crystal of LSCO with $x$=0.03 was grown with the traveling-solvent floating-zone technique [15]. Detailed measurements were carried out on a sample that was cut out along the main crystallographic axes and polished into a bar with dimensions of 2.10x0.44x0.42 mm$^3$, suitable for direct measurements of the in-plane resistance $R_{ab}$. The preparation of the electrical contacts has been described elsewhere [7]. The sample resistance was measured using the standard four-probe ac technique (typically at ~7 Hz) in the Ohmic regime, at $T$ down to 0.05 K realized in a dilution refrigerator.

The sample is located in the SG region of the phase diagram (Fig. 1 inset) at $T \leq T_{SG}$ ~ 7 to 8 K [16]. In that $T$ regime, $R_{ab}$ is insulating with a VRH dependence $R_{ab} = R_0 \exp(T_0/T)^\mu$, $T_0$ = 108 K, $\mu$=1/3, all the way up to $T$=30 K (Fig. 1). ($R_{ab}$ is henceforth denoted by $R$ for simplicity.) The localization length obtained from the VRH fit [9] is $\xi$~90 Å, which is somewhat larger than the average magnetic correlation length ~40 Å at this doping [4, 17]. The most probable hopping distance [18] $r_h(T) \sim \xi(T_0/T)^\mu$ thus spans on the order of ten AF domains at low $T$. Therefore, for the VRH process, the system appears to be uniform. The pMR emerges only deep within the SG phase at $T <$ 1 K $\ll T_{SG}$ [9].

## 3. Results and Discussion

Figure 2 illustrates the response of $R$ upon the application and removal of different $B$ at a fixed, low $T$ in the regime of pMR. Starting with zero-field cooled (ZFC) $R(B$=0), the application of $B$=6 T induces a positive MR. When $B$ is removed, $R(B$=0) is higher than the initial one. Subsequent application and removal of the smaller $B$=4 T does not change the zero-field $R$. Conversely, the sweep to higher $B$=9 T again increases $R(B$=0), which remains unchanged after the final cycling of $B$ between 0 and 2 T. This clearly demonstrates that the system acquires a memory of the highest applied $B$. As long as the processes involve an external $B$ that is smaller than the one previously applied, $R$ will be changed reversibly. These observations are analogous to those for the $c$-axis transport in the low-$T$ pMR regime [6]. The hysteresis and memory are erased by warming the sample up above ~1 K, where pMR vanishes. Importantly, it has been verified that the same results for ZFC $R(B$=0) and MR are obtained also after thermal cycling to $T > T_{SG}$, e.g. to 10 K.

The $c$-axis transport study [6] also revealed that the precise $R_c(T,B$=0) form depends on the cooling conditions. Therefore, ZFC and field-cooled (FC) protocols have been employed here to examine whether the in-plane $R(T)$ also depends on the magnetic history. In the FC protocol, the field is first swept to a particular $B_{FC}$ at a temperature high enough to erase the magnetic memory, i.e. at or above 1 K where pMR and hysteretic effects are absent. After the sweep is stopped, the sample is allowed to cool to the measurement $T$. The field is then swept back to 0 and $R_{FC}(B$=0,$T$) is recorded.

Figure 3 shows that the FC resistance $R_{FC}(B=0)$ is always higher than the zero-field cooled $R_{ZFC}(B=0)$. Repeated cool downs in zero magnetic field yield the values of $R_{ZFC}(B=0)$ to within 1%. On the other hand, the difference between FC and ZFC values is observed to reach up to 80% at the lowest $T$. This difference decreases with increasing $T$, and vanishes at a temperature $T_B$ that grows with $B_{FC}$. While these findings are analogous to those on $R_c$ [6], here the onset of this history-dependent behavior is shifted to considerably lower $T$. For example, for $B_{FC}$=9 T, $T_B$ is about 3 K for the out-of-plane [6] and only ~0.5 K for the in-plane transport (Fig. 3). We recall that the in-plane resistance noise spectroscopy also showed [7] that the emergence of charge glassiness in $CuO_2$ planes takes place at even lower $T$ than for the out-of-plane transport.

Similar memory effects in transport have been observed in other materials with inhomogeneous states, such as manganites [19-22] in the regime of phase separation. The onset of splitting between ZFC and FC data as a function of $T$ was strongly linked to $T_{SG}$ in those materials. In cuprates, magnetoresistive memory has been reported in heavily underdoped $YBa_2Cu_3O_{6+x}$ (YBCO) at low $T$ and attributed to the freezing of the AF domains and their charged boundaries [23]. The onset of the hysteretic MR behavior and of the memory effect in YBCO was also related to the spin-glass transition temperature. It is interesting that the signature of the SG transition in the resistivity of a conventional spin glass (Ag:Mn) has been observed [24] only very recently. In that case, the onset ($T_B$) of the difference between FC and ZFC resistivities also coincides with $T_{SG}$, and it is independent of $B_{FC}$, at least in the experimental $\mu_B B_{FC} \ll k_B T_{SG}$ limit ($\mu_B$ – Bohr magneton). In LSCO, by contrast, $T_B$ decreases with decreasing $B_{FC}$ (Fig. 3 and [6]). Moreover, $T_B \ll T_{SG}$ even for the highest applied field, where $\mu_B B_{FC} \sim k_B T_{SG}$. A similar decrease of $T_B$ as $B \rightarrow 0$ has been observed [9] in the AF (Néel) state of $La_2Cu_{1-x}Li_xO_4$ (Li-LCO) with $x$=0.03. The decoupling of the two temperature scales, $T_B$ and $T_{SG}$, in lightly doped $La_2CuO_4$ suggests that the observed memory effects in the resistance reflect primarily the behavior of charge, not spins.

We also note that these memory effects are not related to the "magnetic shape memory" observed [25] in $x$=0.01 LSCO at room $T$, which was attributed to the alignment of the crystal's orthorhombic $b$ axis along the magnetic field by moving the crystal domain boundaries. In our samples, the applied fields are too small to cause a swapping of the orthorhombic $a$ and $b$ axes. In underdoped to nearly optimum doped $La_2CuO_{4+y}$, it is known that there is a phase separation related to electronic inhomogeneity [26], caused by time-dependent self-organization of interstitial oxygen ions into ordered and disordered nanodomains [27]. However, this effect is extremely unlikely to be related to the memory effects discussed here. Even if some excess oxygen ions exist in our samples, they are mobile only above ~220 K on the time scales of our measurements [27]. In other words, at very low temperatures of our experiment ($T$< 1 K), the ordering of interstitial oxygens would take place on timescales of at least months, probably years [27]. The memory effects observed in the pMR, on the other hand, become more pronounced as $T$ is reduced below 1 K, consistent with the presence of intrinsic dynamic charge inhomogeneities and the underlying charge glass transition as $T \rightarrow 0$. Such charge clusters exist in the regions separating hole-poor AF domains of Cu spins that are frozen into a spin glass.

In $x$=0.03 LSCO and Li-LCO ($T_{SG}$~7 to 8 K [28]), it is known [9] that the ZFC in-plane MR (with $B \parallel ab$) is negligible above 10 K. Below 10 K, the MR becomes negative and increases in magnitude as $T$ is reduced [9, 29, 30]. However, the pMR appears at low fields at $T < T_{SG}$ and dominates the MR over an increasingly large range of $B$ as $T$ is lowered. In contrast to the pMR, the negative component of the MR does not exhibit any hysteretic or memory effects. We note that, at $T<T_{SG}$, the negative MR is observed only for fields $\mu_B B > k_B T_{SG}$. Therefore, it is plausible to attribute the low-$T$ negative MR to the same mechanism that describes the negative MR at higher $T$, namely to an increase of $\xi$ due to the reorientation of the weak FM moments and an effective crossover from primarily 2D to 3D hopping [12]. The reorientation of the weak FM moments also alters the pattern of the AF domains in $CuO_2$ planes, since the direction of weak FM moments is uniquely linked to the phase of the AF order [13, 14, 31]. It is interesting that this effect can be employed to affect the sign of the low-$T$ MR. Figure 4 shows an example of the ZFC positive MR obtained at $T$=0.152 K. However, the MR measured at the same $T$ but obtained after field-cooling in $B_{FC}$=9 T has the opposite sign: it is negative. Since $\mu_B B_{FC} \sim k_B T_{SG}$ was applied at $T > 1$K where the ZFC is negative, it appears that the subsequent cool down to $T \ll 1$ K has frozen the spin configuration that gives rise to this negative MR. More theoretical work is needed, however, to gain a detailed understanding of the interplay of spins and charges in different $T$ regimes and their effect on transport.

**4. Summary**

The memory effects in the low-$T$ positive in-plane magnetoresistance of the insulating, single crystal La$_{1.97}$Sr$_{0.03}$CuO$_4$ have been studied using different experimental protocols. The results strongly suggest that the memory effects reflect primarily the behavior of doped holes.

**Acknowledgements** This work was supported by NSF DMR-0905843, NHMFL via NSF DMR-0654118, MEXT-CT-2006-039047, EURYI, and the National Research Foundation, Singapore.

**Figure 1** The temperature dependence of the zero-field cooled in-plane resistance $R_{ab}$ below 30 K. The solid line is a fit with slope $T_0$. Inset: Schematic phase diagram of hole-doped LSCO.

**Figure 2** The in-plane resistance as a function of time upon the application of $B$=0, 6, 0, 4, 0, 9, 0, 2, 0 T at $T$=0.109 K.

**Figure 3** The difference between FC and ZFC in-plane $R(B=0)$ vs. $T$ for several cooling fields $B_{FC}$. $T_B$ denotes the temperature at which this difference vanishes for a particular cooling field. The dashed lines guide the eye. For $T$=0.115 K and $B_{FC}$=9 T, the results are shown for three different measurements obtained using the same cooling rates.

**Figure 4** Magnetoresistance at $T$=0.152 K after field-cooling ($B_{FC}$=9 T) and zero-field cooling procedures.

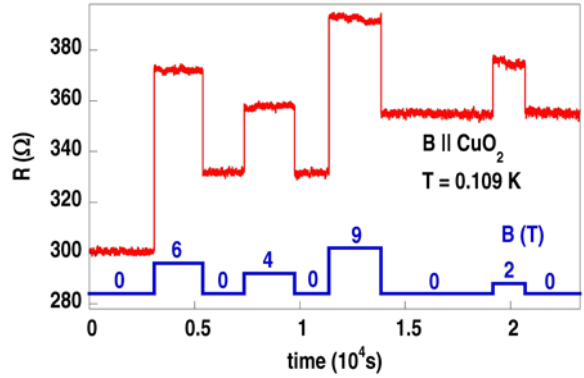

Figure 2

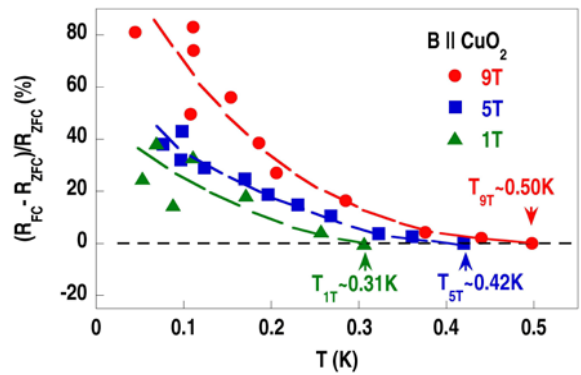

Figure 3

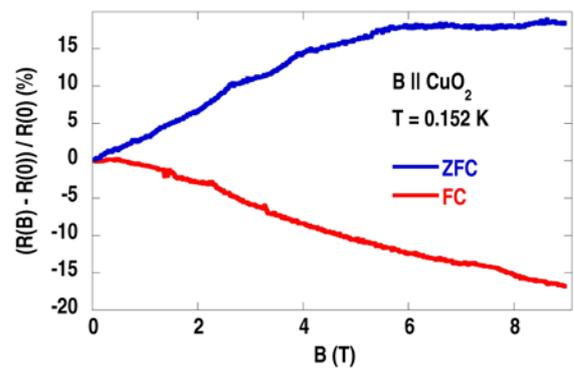

Figure 4

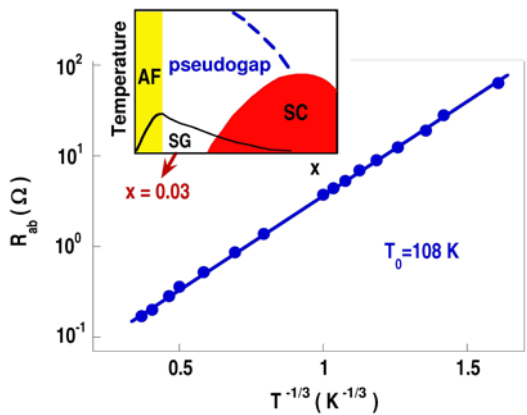

Figure 1